\documentclass{mem}
\usepackage{natbib}
\usepackage{txfonts}
\usepackage{balance}
\usepackage{graphicx}
\usepackage[a4paper,breaklinks,dvipdfm]{hyperref}
\idline{75}{282}
\begin{document}
\def\logg{$\log (g)$~}
\def\Teff{$T_{\rm eff}$~}
\def\FeH{$[Fe/H]$~}
\def\vmicc{$\xi_{t}$}
\def\vmic{$\xi_{t}$~}
\def\msun{$M_{\odot}$~}
\title{The chemical signature of SNIax in the stars of Ursa minor?}

\author {G. Cescutti\inst{1,3} \thanks {email to: cescutti@oats.inaf.it} 
\and  C. Kobayashi\inst{2,3} }
 
\institute{
INAF, Osservatorio Astronomico di Trieste, I-34131 Trieste, Italy
\and Centre for Astrophysics Research, School of Physics, Astronomy and
Mathematics, University of Hertfordshire, College Lane, Hatfield AL10
9AB, UK
\and BRIDGCE UK Network (www.bridgce.net), UK
}

\authorrunning{Cescutti\&Kobayashi }

\titlerunning{Mn in Ursa min: a probe of sub-classes of type Ia SNe?}

\abstract{Recently, a new class of supernovae Ia was discovered: the
  supernovae Iax; the increasing sample of these objects share common
  features as lower maximum-light velocities and typically lower peak
  magnitudes.  In our scenario, the progenitors of the SNe Iax are
  very massive white dwarfs, possibly hybrid C+O+Ne white dwarfs; due
  to the accretion from a binary companion, they reach the
  Chandrasekhar mass and undergo a central carbon deflagration, but
  the deflagration is quenched when it reaches the outer O +Ne
  layer. This class of SNe Ia are expected to be rarer than standard
  SNe Ia and do not affect the chemical evolution in the solar
  neighbourhood; however, they have a short delay time and they could
  influence the evolution of metal-poor systems. Therefore, we have
  included in a stochastic chemical evolution model for the dwarf
  spheroidal galaxy Ursa minor the contribution of SNe Iax.  The model
  predicts a spread in [Mn/Fe] in the ISM medium at low metallicity
  and - at the same time - a decrease of the [alpha/Fe] elements, as
  in the classical time delay model. This is in surprising agreement
  with the observed abundances in stars of Ursa minor and provide a
  strong indication to the origin of this new classes of SNIa.}

\maketitle{}

\section{Introduction}

The scenario leading to a supernova type Ia (SN Ia) explosion is still
under debate. The two most common progenitor scenarios are: the single
degenerate scenario, in which a white dwarf of a mass close to the
Chandrasekhar (Ch) mass accrete mass from a companion - a
 red giant or a main sequence star, and the double degenerate scenario
in which two white dwarfs merge due to the loss of angular momentum.
Both scenarios present pro and contros, but at the present none
of them is able to explain all observational constraints of
SNe Ia \citep[e.g.][]{Maoz14}.   Thanks to the
 observational surveys, a large number of SNe Ia are observed 
  as a luminous and (almost) standard candle, and the SNIa explosions
are fundamental to understand the expansion rate of the Universe bringing up the existence of a
 dark energy, after the dismiss by Albert Einstein several decades
 ago.   On the other hand, the surveys have revealed a small
 variation of the standard candle such as super-luminous or faint
 called Type Iax \citep[e.g.][]{Foley13}.

In general, it is not easy to constrain SN Ia scenarios because the
chemical outcomes are similar.  However, the chemical signature
is relatively clear in the case of Mn, which is one of the few
chemical elements with just a single stable isotope ($^{55}$Mn).
Recently, this chemical element has been investigated in
\citet{Kobayashi15}, where it has been shown how the presence of 
 three channels for the production will promote a different trend in
the chemical evolution for dwarf galaxies, keeping unchanged the trend
in the case of the solar vicinity model. A similar analyses was
performed, but only for the solar vicinity, in \citet{Seitenzahl13},
and the final results was in favor of the presence of two 
scenarios for SNe Ia. Regarding the results for  dwarf galaxies, a
flat and under solar trend for Sagittarius was obtained by
\citet{Cesc08b}, by means of a strong dependence to the metallicity of
the SN Ia yields, keeping again a good agreement for the solar
neighbourhood, the same observational results were obtained for
Sculptor, Fornax, Carina and Sextans in \citet{North12}.   Note
 that, however, such a metallicity effect is not expected for the
 majority of SNe Ia where Mn is synthesised in nuclear statistical
 equilibrium \citep{Kobayashi06,Kobayashi15}.

 In this work we will use the stochastic and inhomogeneous chemical
 evolution models, to study the pollution due to SNe Ia in the
 later stage of a dwarf spheroidal galaxy (dSph), satellite of our
 Galaxy.  At present, not many measurements of Mn are present in
 literature for stars in dSph galaxies.  We decide to compare our
 results with Ursa minor because it is one of the dwarf spheroidal
 with the larger number of stars for which Mn is measured
 \citep{Ural15}, and therefore we decide to apply our modelling to this
 case.

\section{SN Ia scenarios and nucleosynthesis}\label{Nucleos}

In this paper, we consider the same assumptions for sub-luminous (such
as SNe Iax), sub-Ch, and Ch-mass SNe Ia in terms of nucleosynthesis
and rate of the events as in \citet{Kobayashi15}. For comparison, we
present also results based on the SN Ia prescriptions used in
previous papers as \citet{Cesc08b,Spitoni10} based on the original
work presented in \citet{Matteucci1986}.  We decide to call the former
prescriptions ``Herts'' model, whereas the second it is called
``Trieste'' model, from the location of the groups that have developed
them. More details can be found in the papers mentioned above, 
or in the more recent \citet{CescKoba}.

\section {Chemical evolution models for Ursa minor}

We start our analysis on the dSph Ursa minor from the standard chemical
evolution model described in \citet{Ural15}, in particular their 
model C, which takes into account a star formation history based on
the observational constrained by \citet{Carrera02} and infall and winds from the
system able to reproduce the metallicity distribution function
obtained by \citet{Kirby11} \citep[for details, see
Sect. 3][]{Ural15}.
The target of this contribution is however, to investigate the possible
spread produced by a double channel of production of Mn from two
different sources.  Therefore, we have developed a stochastic chemical
evolution model in the same fashion as those implemented in
\citet{Cesc08a,Cesc10,Cescutti13} for the Galactic halo, but with the
specific chemical evolutions of Ursa minor, described above.
As we will see, also in this model we predict a dispersion in the 
first enrichment by SNe II, but also later in the evolution also the
differential production of Mn by the different channel.

\section{Abundances measured in in Ursa minor stars}

We compare our results with the same set of stars shown in
\citet{Ural15}. In this work, the abundance of three stars have been
measured and compared to the abundances collected from other authors
\citep{SCS01,SAI04,Cohen10,Kirby12}. We decide to compare our 
results also with the data coming from other two dSphs 
similar to Ursa minor, Sextans and Carina, using the data available in
\citet{North12}.

\section{Results}

Both scenarios are able to reproduce the
main observed trends within the uncertainties for [Ca/Fe] vs [Fe/H]  
and [Mn/Fe] vs [Fe/H] \citep[e.g.][]{Cesc08b,Kobayashi15,Ural15}
Therefore, at least in first approximation -
the two models are compatible and by means of an
 homogenous model, it was not possible to distinguish between them.
Therefore, in this work we use a stochastic model to outline a
different conclusion, for the Mn case.  

\begin{figure}[ht!]
\includegraphics[width=75mm]{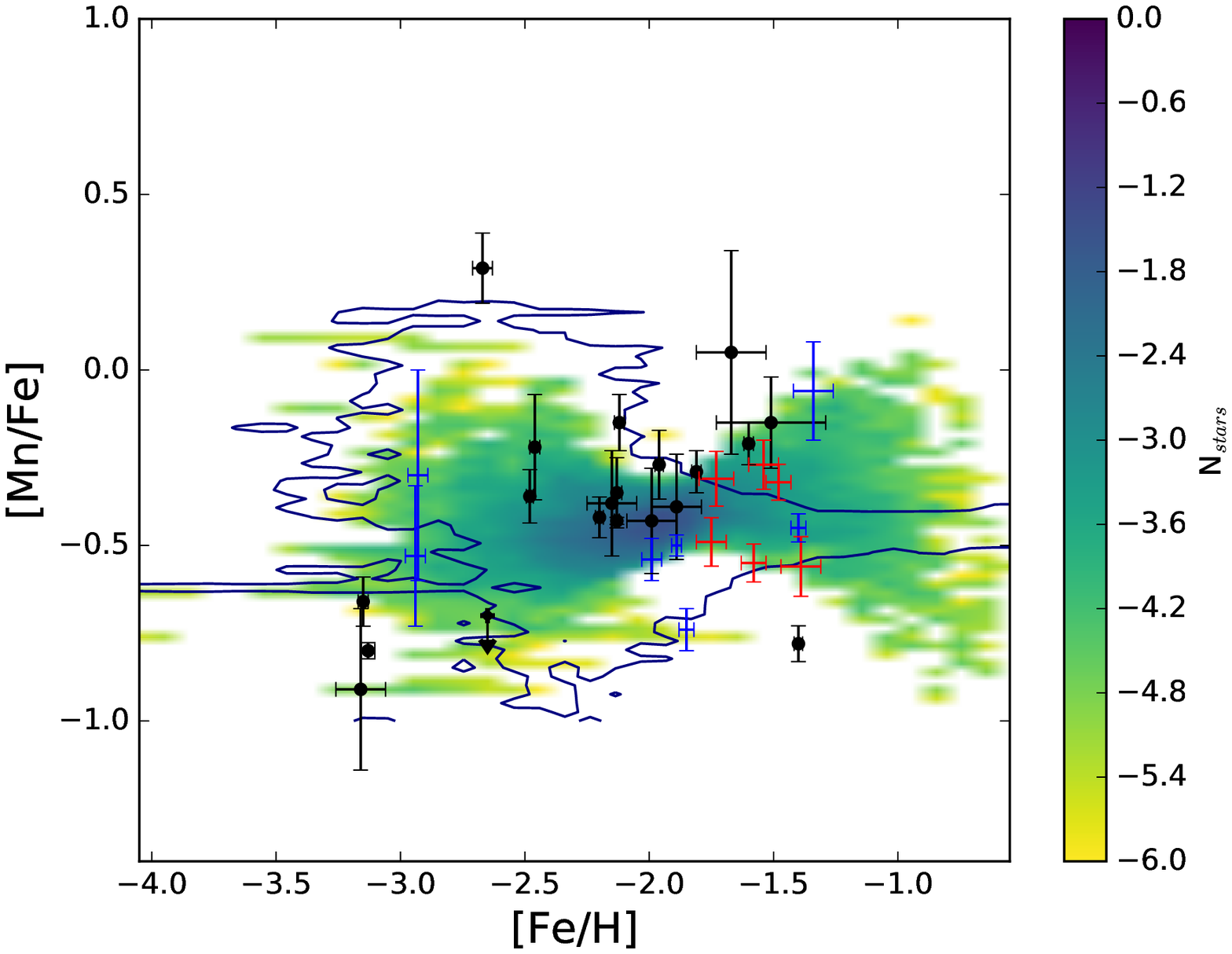}
\caption{ [Mn/Fe] vs [Fe/H], the data in black represent the
  abundances measured in stars belonging to the dSph Ursa minor presented in
  \citet{Ural15}; in red we present the data for the dSph Carina from
  \citet{North12} and
  in blue those for the dSph Sextans again from from \citet{North12}
  and \citet{TJH10}. The colour-coded surface density
  plot presents the density of long living stars for the Herts model,
  the contour line the same results for the Trieste model.
}\label{fig5}
\end{figure}

In the Trieste model (Fig. \ref{fig5}, the contour plot) the strong dependence of the
[Mn/Fe] to the stellar mass in massive stars, produce a large spread
for [Fe/H]$<-$2.  On the other side, 
at [Fe/H]$>-$2, when the SNe Ia start to play an important role, the
spread for the Trieste model in the [Mn/Fe] vs [Fe/H] space is
decreased, due to the approximately constant enrichment of Mn and Fe
from the SNe Ia. In this region, the model does not agree with any of
the abundances measured in the 4 stars of Ursa minor (black symbols).

The  Herts model with its stochastic
results  shown in Fig.\ref{fig5} displays a butterfly shape
distribution with  remarkable differences compared to  the Trieste model.
Again the bulk of the data at [Fe/H]$\sim-$2 are within the
prediction of the model; moreover at lower metallicities most
of the measured stars are in good agreement with the model, and in this
case the three stars at [Fe/H]$<-2.5$ and [Mn/Fe]$<-$0.5
are better within the limits of the probability predicted by the
model. 
The most striking difference is on the high metallicity tail. In this
region, at [Fe/H]$>-2$, the model produces again a spread (the
right wing of the butterfly!). This is due to the onset of the 
two SN Ia channels that start to enrich the ISM. Producing a difference
[Mn/Fe] ratio, they create a spread in the model results: regions
polluted by the SNe Iax tend toward solar [Mn/Fe] ratios, the contrary
for the sub-Ch SNe Ia.

Comparing the data we have for this galaxy, it clearly appears that the
Herts model is the one that more closely approximates what is 
displayed by  the stars of Ursa minor. In fact three stars 
have a [Mn/Fe]$>-$0.5 and one instead have a lower value,
presenting therefore a spread. At this stage, the number of data are not yet statistically 
enough to ensure that our prediction is firm, and future observational
campaign to measure more spectra of stars in Ursa minor is encouraged.
Moreover, we underline that it will be vital for this project to measure
not the most extreme metal poor tail, as commonly happens, but
the opposite, the metal rich end, in order to disentangle this problem.
As mentioned in the introduction, not many measurements of Mn are
present in literature for stars in dSph galaxies. In this respect a
significant work has been carried out in the paper \citet{North12},
where the measurement of four dSph galaxies are presented: Sculptor,
Fornax, Carina and Sextans. 
\vspace{-0.5cm}
\vspace{-0.25cm}
For this reason, we
include also the stars measured for Carina and Sextans,
two dSphs with stellar mass similar to Ursa minor,  in total 13 more data points
in the plot. Moreover eight of these data sit at [Fe/H]$>-1.8$,
increasing significantly the only 4 data available for Ursa minor in
this metallicity range, where the new Herts model predicts a spread in [Mn/Fe].
Although the data belong to two different dSphs, it is still
encouraging to see that the model is in excellent agreement with them;
in fact the data for Sextans for example show a remarkable spread
 of more than 0.5 dex.
Therefore, we encourage more investigations to establish the presence
 of this spread in the [Mn/Fe] vs [Fe/H] space in other satellites of
 the Milky Way, in particular in the faint classical dSphs as Ursa
 minor, Sextans, Carina and Draco.  

\section{Conclusions}

We present new results for the chemical evolution of the [Mn/Fe] in
the dwarf spheroidal galaxy Ursa minor with two different
prescriptions for the SNe Ia. These two prescriptions that we call
Herts and Trieste, present these differences: in the Trieste model we
allow to explode only a single channel of SNe Ia, the single
degenerate with a deflagration; in the Herts model we have two
different channel, one is a sub-Ch channel, with a double
detonation, the other is a special case of single
degenerate, originated from a relatively massive primary star, producing a
relatively weak deflagration (SN Iax channel).  These two channels produce on
the average the same amount of Mn as the the single channel in the Trieste
model.  We show that in the framework of an homogenous chemical
evolution model, both Herts and Trieste prescriptions are compatible
with the data available for Mn in this dSph.  On the other hand, in
the stochastic framework, the results are quite different and the data
seem to favour the Herts model, and therefore, the presence of two
channels for SNe Ia at low metallicity, in addition to normal SNe Ia
at higher metallicities.  
Clearly, also more data are important to raise firmer
conclusions, and we encourage to test our thesis by analysing the stars belonging to 
the tail at higher metallicities for this class of dSphs, rather than
focusing the most extreme metal-poor component.

\begin{acknowledgements}
 This work was supported by the Science and Technology Facilities
  Council ST/M000958/1 for the BRIDGCE consortium grant.
G.C. acknowledges financial support
from the European Union Horizon 2020 research and
innovation programme under the Marie Sk\l odowska-Curie
grant agreement No. 664931.
\end{acknowledgements}

\bibliographystyle{aa}
\bibliography{spectro}

\end{document}